\documentstyle[twocolumn,prl,aps,subfigure,psfig,subeqnar]{revtex}

\newcommand{\MM}{\,\mbox{\bf M}}

\newcommand{\JM}{\,\mbox{\bf J}}

\newcommand{\tr}{\,\mbox{\rm Tr}}
\newcommand{\Lp}{{\bf L_+}}
\newcommand{\Lm}{{\bf L_-}}

\newcommand{\sn}{\mbox{\rm sn}}

\newtheorem{thm}{Theorem}
\newtheorem{corr}{Corollary}

\psfull

\begin{document}
\title{Instabilities in the Bogoliubov Spectrum of 
a condensate in a 1-D periodic potential}

\author{Jared C. Bronski$^{1}$, Zoi Rapti$^{2}$\cite{byline}\\}
\address{$^{1}$Department of Mathematics, University of Illinois 
         Urbana-Champaign, Urbana, IL 61801, USA\\}
\address{$^{2}$School of Mathematics, Institute for Advanced Study, 
         Princeton, NJ 08540, USA\\}

\maketitle

\date{\today}

\begin{abstract}

We study the stability of standing wave solutions to a 
one-dimensional Gross-Pitaevsky equation with a periodic 
potential. 
We use some simple complex analysis and the Hamiltonian structure 
of the problem to give a simple rigorous criterion which guarantees 
the existence of non-real spectrum, which corresponds to exponential 
instability of the standing wave solution. This criterion can be stated 
simply in terms of the spectrum of one of these self-adjoint operators. 
When the standing wave has small amplitude this criterion simplifies 
further, and agrees with arguments based on the effective mass in the 
periodic potential.

\end{abstract}


\maketitle
The possibility of creating cold atom condensates in optical lattices 
has made it possible to observe both linear phenomena, such as Bloch 
oscillations\cite{BO} as well as genuinely nonlinear phenomena such as soliton 
generation\cite{SMBFMMCA} and an insulating transition\cite{SMSKE}. 
While a great deal of effort has gone into the construction of 
exact or approximate solutions the question of stability is
foremost, since it dictates what states can actually be observed,
as well as governing the evolution of small disturbances to these
states. Recent papers of Wu and Niu\cite{WN0,WN1} and 
Burger et.al. \cite{BCFMICT} have helped elucidate the 
role of the dynamic and Landau instabilities.
These dynamic instabilities have also been realized 
experimentally\cite{FSLMSFI}.
Most of the current theoretical stability work is either 
numerical or relies on either special properties of some exact 
solution\cite{us1,us2,ABD}, or some 
approximation, most notably a two-level approximation\cite{WN1}, the  
tight binding approximation\cite{WN1,STKB,MST},
weak nonlinearity\cite{KS}, or long wavelength\cite{TZ}. In this paper we 
give a simple, rigorous, easily 
checked sufficient condition for the existence of exponential 
instabilities. In the case of a weak nonlinearity this condition 
reduces to a physically natural condition in terms of the effective mass
of the lattice.

Our starting point is 
the 1-D Gross-Pitaevsky equation with a periodic potential,
\[
i\psi_t = -\frac{1}{2} \psi_{xx} \pm |\psi|^2 \psi + V(x) \psi ~~~~ 
V(x+1)=V(x)\]
with a standing wave solution of the form $\psi(x,t) = \exp(-i\omega t) 
\phi(x)$, where the standing wave
profile $\phi$ is real and  either periodic $\phi(x+1) = \phi(x)$ or 
antiperiodic $\phi(x+1) = -\phi(x)$. The linearized equation governing 
small disturbances is 
\begin{eqnarray*}
i u_t &=&  \left(-\frac{1}{2} \partial_{xx} + V(x) + 
|\phi|^2(x) - 
\omega\right) v  = \Lp v \\
i v_t &=& \left( -\frac{1}{2} \partial_{xx} + V(x) + 
3|\phi|^2(x) - \omega \right) u = \Lm u  
\end{eqnarray*}
where $u,v$ are the real and imaginary parts of the complex disturbance.
Upon taking a Fourier transform in time this takes the form of the well-known 
Bogoliubov equation
\begin{eqnarray*}
\mu u &=& \Lp v \\
\mu v &=& \Lm u.
\end{eqnarray*}
Because of the Hamiltonian structure of the problem, the above eigenvalue 
problem can be written in vector form
\[
\vec v_x = {\bf J} {\bf H}(x,\mu) \vec v
\]
with $\vec v = (u, v, u_x,v_x)^t,$ the matrix ${\bf H}={\bf H}^\dagger$ 
is Hermitian, and 
the matrix ${\bf J}$ is the canonical Hamiltonian form ${\bf J } = 
\left(\begin{array}{cc} 0 & {\bf I} \\ -{\bf I} & 0 \end{array}\right).$
The standard results of Floquet theory guarantee that 
$\mu$ is in the spectrum of the operator above if and only if 
the period map ${\bf M}(\mu)$ has eigenvalues on the unit circle, where 
the period map ${\bf M}(\mu)$ is defined by 
\begin{eqnarray}
{\bf U}_x &=& {\bf J} {\bf H(x,\mu)} {\bf U} ~~~ {\bf U}(0,\mu) = {\bf I} \\
{\bf M}(\mu) &=& {\bf U}(1,\mu)
\end{eqnarray}
In the second order case, which arises in the classical Floquet-Bloch theory, 
the location of the eigenvalues of the period map are determined 
entirely by the trace of the ($2\times2$) matrix ${\bf M}$, there being two 
eigenvalues on the unit circle if $\tr({\bf M}) \in (-2,2)$ and 
two eigenvalues off of the unit circle if  $\tr({\bf M}) \notin [-2,2]$.
The case of the Bogoliubov operator is somewhat more complicated, since it 
consists of coupled second order problems, but 
can be treated in much the same way due to the Hamiltonian structure\cite{YS}.
First we note that ${\bf U}$ and thus ${\bf M}$ satisfy the 
relation 
\[
{\bf U}^t(x,\mu) {\bf J} {\bf U}(x,\mu) = \JM,
\]
a result known as the Poincare-Liouville theorem. 
From this it follows that ${\bf M}$ is similar to its inverse transpose. 
This implies that the eigenvalues are invariant under inversion with respect
to the unit circle, which in turn implies that the characteristic 
polynomial of ${\bf M}$ has the following form:
\begin{eqnarray}
\det({\bf M} - \lambda {\bf I}) &=& 1 + a \lambda + b \lambda^2 + a 
\lambda^3 + \lambda^4 \\
a &=& -\tr({\bf M}) \\
b &=& \frac{1}{2} \left(\tr({\bf M})^2 - \tr({\bf M^2}) \right).
\end{eqnarray}
Because of the special form the above, the quartic admits an explicit 
factorization into quadratics,
\[
\det({\bf M} - \lambda {\bf I}) = (1 - K_+(\mu) \lambda + \lambda^2) 
(1 - K_-(\mu) \lambda + \lambda^2),
\]
where the quantities $K_\pm$ are given by 
\[
K_\pm(\mu) = \frac{\tr({\bf M}) \pm \sqrt{2\tr{\MM^2}(\mu)-
(\tr{\MM}(\mu))^2+8}}{2}.
\]
It then follows that the period map for the Bogoliubov operator 
has two eigenvalues on the unit circle if $K_+$ is real
and $K_+\in (-2,2)$, and another two on the unit circle if $K_-$ is real
and $K_-\in (-2,2)$. It can be shown using complex analysis that 
the spectrum of the Bogoliubov operator consists of a union of 
piecewise smooth arcs in the complex plane, where the endpoints of the 
arcs are either band edges $K_\pm(\mu)=\pm 2$, critical points 
$K_\pm^\prime(\mu) = 0$, or branch points $K_+(\mu)=K_-(\mu)$.
It is worth noting that a generic quartic does not admit a 
factorization into quadratics of the above form. It is only because 
of the Hamiltonian nature of the problem that this occurs. For a fourth order 
problem without this symmetry it is necessary to use the general
solution to the quartic, and a calculation analogous to the present 
would be exceedingly difficult.

We are primarily interested in instabilities of modulational type, where the 
Bogoliubov operator has spectrum off of the real axis in a neighborhood of 
$\mu=0$. It is also possible for the operator to exhibit sideband type
instabilities, but these instabilities are more difficult to detect. 
We will consider these in a future paper. To derive a criterion for 
a modulational instability, we note the following facts
\begin{itemize}
\item $\tr\left(\MM^2(\mu)\right),\tr\left(\MM(\mu)\right)^2$ are even,
 analytic functions of a complex $\mu$
\item $K_\pm(\mu)$ are even and analytic in a neighborhood of 
$\mu=0$ if $K_+(0)\neq K_-(0)$.
\item $\Lm$ (considered as a periodic Schr\"odinger operator) has a 
band edge at $\mu=0$.
\end{itemize}

The first follows from the fact that the Bogoliubov operator is invariant 
under the transformation $u \rightarrow -u, v \rightarrow v, \mu 
\rightarrow -\mu$. This implies that 
the monodromy matrix $\MM(\mu)$ satisfies $\MM(\mu) = {\bf V} \MM(-\mu) 
{\bf V}^t$ with ${\bf V}$ orthogonal, which implies that 
$\tr\left(\MM(\mu)\right) =\tr\left(\MM(-\mu)\right)$ and likewise for 
$\tr\left(\MM^2(\mu)\right).$ This symmetry reflects the symmetry
between left and right propagating waves. This symmetry is not present 
if one considers the linearization about topological-type standing 
waves which possess a non-trivial quasi-momentum, and the analysis 
is much more difficult.  Standard 
arguments from the theory of ordinary differential equations show that 
 $\tr(\MM(\mu))$ is an analytic (matrix-valued) function of a complex 
$\mu$. It follows 
from the formula for $K_\pm$ that, away from the branch points where 
$(K_+(\mu)-K_-(\mu))^2= 2\tr{\MM^2}(\mu)-\left(\tr{\MM}(\mu)\right)^2+8=0$, 
the Floquet discriminants $K_\pm(\mu)$ are analytic functions. The final 
fact follows from the ${\cal U} (1)$ symmetry of the Gross-Pitaevsky equation 
$\psi \rightarrow \psi e^{i \theta}$, which (via Noether's theorem)
implies that $L_-\phi = 0$. Since $\phi$ is either periodic or antiperiodic 
this implies that zero energy is a band edge for $\Lm$.

The main result of this paper is the following:
\begin{thm}
Suppose that $\phi$ is a standing wave solution to the GP equation 
for which the $\Lp$ operator is in the interior of a band. Then 
this solution to the GP equation is exponentially unstable. 
\end{thm}

{\it Proof:} When $\mu=0$ the equations for $u$ and $v$ decouple, and 
the period map for the GP equation is block diagonal, with one $2\times2$ 
block (denoted by ${\bf m}_\Lp$ corresponding to $\Lp$,  and the other 
block (${\bf m}_\Lm$) corresponding to $\Lm$. If $\mu = 0$ is not a 
band edge for $\Lp$ it follows that $(K_+(0)-K_-(0))^2 = 
\left(\tr({\bf m}_\Lm) - \tr({\bf m}_\Lp)\right)^2 \neq 0$, so 
the Floquet discriminants are analytic in a neighborhood of $\mu=0$.
Since $K_\pm$ 
are even functions it follows that both are real on some interval of the 
imaginary axis. If $\mu=0$ is in the interior of a band for $L_+$ 
it follows that $K_+(0) \in (-2,2)$, and so there is some interval 
along the imaginary axis on which $K_+$ is real and $\in (-2,2)$. 
Thus the GP equation has a band of spectrum along the imaginary axis. 

An important special case of this is given by the following:
\begin{corr}
Small amplitude solutions of the focusing GP equation corresponding to 
lower band edges are modulationally unstable, small amplitude solutions 
of the defocusing GP equation corresponding to upper band edges are 
modulationally unstable. 
\end{corr}

{\it Proof:} The proof is perturbative, considering $\Lp$ as a perturbation
of $\Lm$. Since $\Lp = \Lm \pm 2 |\phi|^2 $ the perturbation is strictly 
positive (resp. negative) and it follows that the band edges of the $\Lp$
and $\Lm$ operators satisfy $\mu_n(\Lp) \ge \mu_n(\Lm)$ (resp. $\mu_n(\Lp) 
\le \mu_n(\Lm)$).  
Since the unperturbed operator $\Lm$ 
has a band edge at $\mu=0$ it follows that, for $||\phi||_\infty <\!\!< 1$, 
$\mu=0$ will be in the interior of a band of $\Lp$ for the focusing case 
and a lower band edge, or defocusing case and an upper band edge. 
From the above result this guarantees the existence of an instability. 

This result provides a rigorous justification of the intuition provided 
by the effective mass theories, such as those proposed by Konotop and 
Salerno\cite{KS} and Taylor and Zaremba\cite{TZ} where the effective mass 
is given by 
the curvature of the dispersion relation. Since this curvature has 
the same sign as the bare mass at lower band edges and the opposite sign
at upper band edges this kind of argument leads to the same conclusion. 
Another way to look at this result is that in either of the above cases 
(lower band edge with a focusing nonlinearity or upper band edge with 
a defocusing nonlinearity) the periodic solution is unstable to formation 
of a gap soliton. The obvious advantage of the more general criterion is 
that it applies for strong nonlinearity, when the band-gap structure 
of ${\bf L_\pm}$ can differ considerably from the linear problem, and 
effective mass arguments based on the linear problem can no longer be
expected to hold.

\begin{figure}
\centerline{\psfig{figure=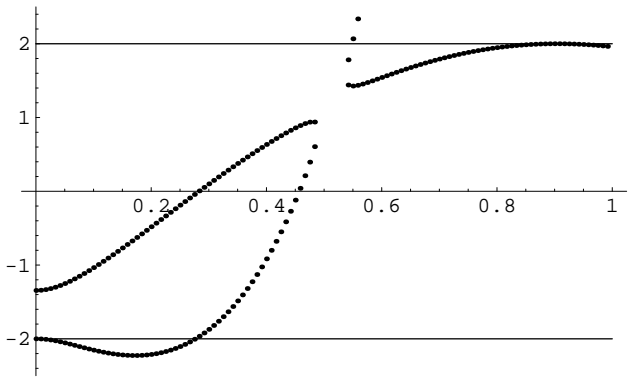,width=83mm}}
\begin{center}
\begin{minipage}{83mm}
\caption{Plot of the Floquet discriminants $K_\pm(\mu)$ for $\mu$ along the 
real axis for an $\sn(x,k)$ type solution to the focusing GP equation with
an elliptic function potential. Note that $K_+(0) \in(-2,2)$}
\label{fig:spectrum1}
\end{minipage}
\end{center}
\end{figure}

\begin{figure}
\centerline{\psfig{figure=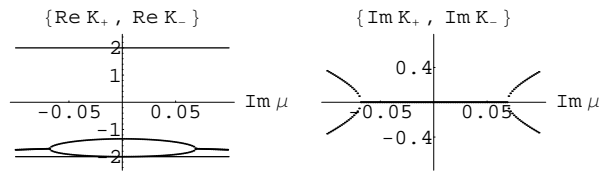,width=83mm}}
\begin{center}
\begin{minipage}{83mm}
\caption{Plot of the Floquet discriminants $K_\pm(\mu)$ for $\mu$ along the 
imaginary axis for an $\sn(x,k)$ type solution to the focusing GP equation 
with an elliptic function potential. Note that $K_\pm(\mu)$ are real 
in an interval of the imaginary $\mu$ axis containing the origin. 
The arc of spectrum terminates at a pair of branch points
at $\mu \approx \pm .07 i $}
\label{fig:spectrum2}
\end{minipage}
\end{center}
\end{figure}

We have conducted some numerical experiments to illustrate these results. 
For our basic model we take the one dimensional Gross-Pitaevsky equation 
with a Jacobi elliptic function potential:
\[
i\psi_t = -\frac{1}{2} \psi_{xx} \pm |\psi|^2 \psi + V_0 \sn^2(x,k) \psi.
\]
As is shown in previous work\cite{us1} this equation has a family of 
elliptic function solutions. In the first experiment we take the 
focusing sign of the nonlinearity. In this case we have an exact solution 
\[
\psi(x,t) = \sqrt{-(V_0 + k^2)} \sn(x,k) \exp(-i\omega t) ~~~~ V_0 \le -k^2
\]
which represents a solution bifurcating from the third band edge - the 
lower band edge of the second band. The corollary implies that for 
$|V_0 + k^2| <\!\!< 1$ there exists a band of spectrum along the imaginary 
axis. Figure \ref{fig:spectrum1} shows a plot of the Floquet discriminants
$K_\pm(\mu)$ along the real $\mu$ axis for $k^2 = .5, V_0 = -.6$. 
Note that at $K_-(0)=-2$, 
via Noether's theorem and the fact that the third band edge represents 
an antiperiodic eigenfunction. The other Floquet discriminant satisfies 
$K_+(0) \approx -1.4$, and is thus in the interior of a band. From the 
first theorem it follows that there is a band of spectrum along the 
imaginary $\mu$ axis. Figure \ref{fig:spectrum2} shows a plot of 
$K_\pm(\mu)$ along the imaginary axis. Both Floquet discriminants 
are real and $\in (-2,2)$ near the origin, implying that the 
Bogoliubov eigenvalue problem has a band of spectrum along the imaginary
axis. This corresponds to an exponential instability in the 
evolution of small disturbances to this standing wave solution. 
 
In summary, by using the Hamiltonian structure and symmetries of the 
Gross-Pitaevsky equation with a periodic potential, we have presented 
a sufficient condition for the exponential instabilities of standing 
wave solutions.

It is our hope, that this criterion will be a complement to the numerous 
already known solutions in studying the Gross-Pitaevsky equation, and 
condensates in optical lattices.


\begin{thebibliography}{10}
\bibitem[*]{byline} to whom correspondence should be addressed
\end{thebibliography}
\end{document}